\documentclass[aps,prd,groupedaddress,preprint,eqsecnum,nofootinbib]{revtex4}
\usepackage{graphicx,epsf,amssymb,amsbsy,amsfonts,amssymb,amsmath}

\def\be{\begin{equation}}
\def\ee{\end{equation}}

\def\bg{\bar{g}}
\def\beq{\begin{eqnarray}}\def\eeq{\end{eqnarray}}
\def\ba#1\ea{\begin{align}#1\end{align}}
\def\bg#1\eg{\begin{gather}#1\end{gather}}
\def\bm#1\em{\begin{multline}#1\end{multline}}
\def\bmd#1\emd{\begin{multlined}#1\end{multlined}}

\def\({\left(}
\def\){\right)}
\def\[{\left[}
\def\]{\right]}

\pdfoutput=1

\begin{document}
\hfuzz 12pt
\title{Kinematics of Horizon and Singularity and IR/UV Mixing in AdS}
\author{Shamik Banerjee}
\affiliation{Institute of Physics, \\ Sachivalaya Marg, Bhubaneshwar, India-751005 \\ and \\ Homi Bhabha National Institute, Anushakti Nagar, Mumbai, India-400085}

\email{banerjeeshamik.phy@gmail.com}
\begin{abstract}

In [arXiv:1512.02232 [hep-th]] it was argued, based on the construction of a holographic c-function, that the curvature singularity of a black brane can be thought of as a trivial IR fixed point. This is dual to the gapped nature of the thermal state in the IR. So one can say that by taking one kind of low energy limit in the thermal CFT we are probing the near-singularity region. But there is another more conventional low energy limit which corresponds to probing the near-horizon region. Now, instead of one, if we think in terms of these \textit{two} low-energy limits and take into account the fact that in AdS-CFT the only observables are CFT correlators then we can get a completely different interpretation of the curvature singularity. In a nutshell, the very long wavelength degrees of freedom in the thermal CFT carry information of \textit{both} the near-horizon \textit{and} the near-singularity regions, but, the field theory observer \textit{cannot}, in principle, disentangle the information of the near-horizon region from the information of the near-singularity region using the \textit{the thermal CFT correlators}. This can be interpreted as a very specific form of \textit{holographic} "non-locality" in a black hole background which relates the "inside and the outside". We argue in the paper that owing to this "non-locality", the space-like curvature singularity along with its problems, which are all \textit{local} in nature, completely \textit{disappear from the theory} or get \textit{dissolved}. But, the same "non-locality" now tells us that some of the "$e^{-S}$-effects" that one finds, for example, in the late time thermal two-point function, can be thought of as carrying \textit{complete} information about "Planck-scale effects near the singularity". From the local EFT point of view this may be called "UV-IR-mixing" which is caused by the "non-locality".  We also discuss its close connection to black hole complementarity.

\end{abstract}

\maketitle
\tableofcontents


\section{A Tale of Two Infrareds}

\subsection{RG-flow in the infalling-frame and Classical Black Hole Singularity}

In this paper we consider black brane in AdS formed by collapsing matter which is dual to an approximately thermal state of the boundary CFT living on Minkowski space-time. We are interested in sufficiently late time when the bulk settles down to the static AdS-Schwarschild black brane to a very good approximation. 

In \cite{Banerjee:2015coc} it was suggested that from the holographic RG point of view the curvature singularity can be thought of as a trivial IR fixed point and the high temperature (IR) expansion of the renormalized logarithmic negativity of the dual CFT is a way to probe the region near the classical singularity. The suggestion was based on the following observations : 

\begin{figure}[htbp]
\begin{center}
  \includegraphics[width=13cm]{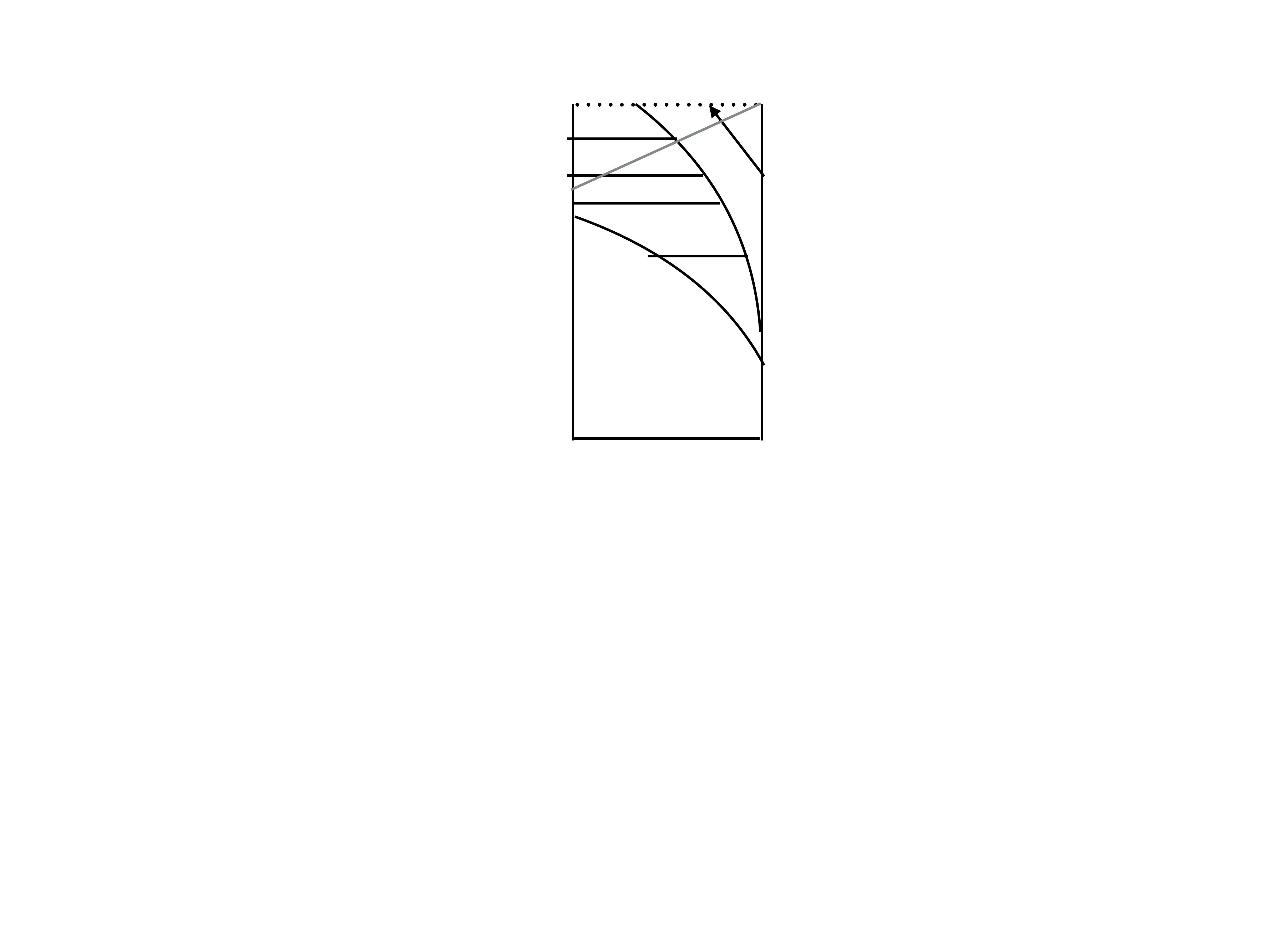}
\end{center}
\vspace{-12em}
\caption{The figure shows a black brane in AdS formed from collapse. The arrow denotes the future bulk light-cone which starts at a boundary point at sufficiently late time and ends at the curvature singularity.}
\end{figure} 

1) In AdS-CFT correspondence \cite{Maldacena:1997re,Witten:1998qj,Gubser:1998bc} radial direction in the bulk is identified with the field theory energy scale \cite{Susskind:1998dq}. So moving along the radial direction can be thought of as renormalization group (RG) flow in the dual field theory. The irreversibility of the RG flow is encoded in the $c$-theorem \cite{Zam} and a holographic $c$-function can be constructed in certain geometries \cite{Akhmedov:1998vf}. In \cite{sb, Banerjee:2015uaa} it was shown that if we consider a \textit{domain-wall geometry} then the holographic $c$-function can be given an entropic interpretation in the following way : Consider a point on the boundary of the AdS and draw the future bulk light-cone of that point. Now construct space-like cross sections of the future light-cone and assign Bekenstein-Hawking entropy to each cross section. The cross sections are non-compact in this case and so the right quantity to consider is the entropy density. It turns out that this entropy density \textit{decreases monotonically} as we move deeper into the bulk along the future directed null geodesic generators of the light-cone. The numerical value of the density coincides with the UV and the IR central charges of the dual field theory when evaluated in the UV region and the IR region of the domain-wall geometry. Hence this entropy density is a holographic $c$-function. The future bulk light-cone of a boundary point is also known as the past causal horizon and the monotonic decrease of the holographic $c$-function also follows from the second law of causal horizon thermodynamics \cite{Jacobson:2003wv}. So the dual of the field theory $c$-theorem may be identified with the second law of causal horizon thermodynamics in the bulk. 
 
 This has the virtue of being a covariant prescription for constructing a holographic $c$-function and can be immediately generalized to other geometries. The challenge here is of course to interpret these generalized $c$-functions in the dual field theory.

 2) We can now apply this to an $AdS_5$ black brane \cite{Banerjee:2015coc}. Black brane is not a relevant deformation of the field theory. But due to finite temperature scale invariance is broken and various quantities in field theory show interesting scale variation. We call this RG-flow. 
 
 In this case the future bulk light-cone of a boundary point ends in the curvature singularity (See Fig-1). Once again we can construct a holographic $c$-function as the Bekenstein-Hawking entropy density on the space-like slices of the light-cone. It turns out that the holographic $c$-function has the value $a_{UV}$ near the boundary of AdS and then it \textit{monotonically} decreases to \textit{zero} at the curvature singularity. Therefore \textit{curvature singularity can be thought of as a trivial IR fixed point of a gapped system} \cite{Banerjee:2015coc}. 
 
 The gapped system is not difficult to identify in the dual field theory. The thermal state which is dual to the black brane behaves like a gapped system in many respects. For example, equal time two point correlation function at finite temperature decays exponentially if the separation between the insertion points is much larger than the inverse temperature. So inverse temperature acts as finite correlation length. Another manifestation of the gapped nature is the absence of any \textit{long-range} quantum entanglement at finite temperature. This property turns out to be very useful for constructing potential candidate for the $c$-function in thermal CFT. The field theory $c$-function should have all the properties of the holographic $c$-function that we have constructed. So the UV value should be given by the central charge of the CFT and then it should decrease monotonically to \textit{zero} in the IR. In \cite{Banerjee:2015coc} it was pointed out, based on the calculations of \cite{Calabrese:2014yza}, that in a two dimensional thermal CFT, the UV value of \textit{renormalized logarithmic negativity} is given by the central charge of the CFT and the IR value is \textit{zero}. So in two dimensions the renormalized logarithmic entanglement negativity is a \textit{potential} candidate for the $c$-function. The vanishing of logarithmic negativity in the IR is just a reflection of the fact that it is an \textit{entanglement measure} for mixed states \cite{Vidal:2002zz} and so can detect the absence quantum entanglement in the IR. In higher dimensions also we expect the renormalized logarithmic negativity to show the same \textit{asymptotic} behavior based on this physical consideration. What is not clear is whether the decrease from the UV to the IR is monotonic or not although there are some numerical evidence that this may indeed decrease \cite{Calabrese:2014yza}.
 
 Another way to think about this is the following. Suppose we define an "effective central charge" $c_{eff}$ in the thermal CFT using say the renormalized logarithmic negativity. In the UV the value of $c_{eff}$ is given by the central charge $c$ of the CFT whereas in the IR it decreases to zero. In the gravity approximation in AdS$_{d+1}$,  $c\sim (\frac{L_{AdS}}{L_pl})^{d-1} \sim O(N^2)$. This is the UV value of $c_{eff}$. Now as we go to longer distances in the CFT,  $c_{eff}$ decreases and at some point in the deep IR,  $c_{eff} \sim (\frac{L_{eff}}{L_pl})^{d-1} \sim O(1)$. So the IR of the thermal state, \textit{as probed by this $c$-function}, is dual to a strongly coupled bulk region. In the black brane geometry it is natural to identify this as the region near the curvature singularity. 
 
 Another evidence for the \textit{gapped nature of the black hole interior} comes from the tensor network construction of the thermofield double state \cite{Hartman:2013qma} which is dual to an eternal black hole in AdS. 
 
Now the above description, based on the construction of holographic c-function or its potential field theory candidate at finite temperature, suggests that the IR or very long wavelength degrees of freedom of the CFT at finite temperature encodes information about the region near the classical singularity. One can probe this by studying, for example, the RG-flow of   \textit{quantum entanglement} in the thermal state. Lack of long range quantum entanglement in the thermal state is manifested in the presence of the curvature singularity in the bulk. This is consistent with the well-known scale-radius duality in AdS-CFT \cite{Susskind:1998dq} and we can think of this as \textit{scale-radius duality in the infalling frame where the IR of the field theory is identified with the curvature singularity.} 

The goal of this paper is to argue that this description of the classical singularity is essentially "classical". In quantum gravity, interpretation of curvature singularity seems to be drastically different and the surprising fact is that the \textit{presence of the horizon as a region of high redshift forces this interpretation upon us.}

\subsection{RG-flow in the asymptotic frame and Black Hole Horizon}

From the point of view of the asymptotic observer space-time geometry ends at the horizon of the black hole. The asymptotic observer's time or the Schwarschild time can be identified with the global (Minkowski) time of the field theory and no bulk object can cross the horizon in finite Schwarschild time. Therefore the standard time evolution in the field theory does not describe horizon crossing. 

Scale-radius duality is an important component of AdS/CFT . It is well understood that in a black brane geometry, the near-horizon region corresponds to the IR of the dual field theory and the Schwarschild radial coordinate can be identified with the energy (RG) scale. This can be thought of as the \textit{scale-radius duality in the asymptotic frame}. The most prominent reason for this is that the near-horizon region is a region of high redshift. So a bulk object placed very close to the horizon has very little field theory energy. Moreover a boundary excitation with size of the order of the thermal scale or bigger can be thought of as residing in the near horizon region in the bulk as has been argued for example in \cite{Banks:1998dd,Balasubramanian:1998de}. There is also strong indication coming from the holographic Wilsonian RG flow approach to various low energy phenomena in the dual field theory \cite{Heemskerk:2010hk}. Now, this description of the scale-radius duality where the near horizon region appears in the IR is the one appropriate for an asymptotic observer. This should be contrasted with the \textit{scale-radius duality in the infalling frame} in which the curvature singularity appears in the IR. 

Let us now discuss a simple example which is helpful for the purpose of visualisation. 
\

Consider RG flow of the thermal state in the CFT. A well known property of a thermal state is the absence of long range correlation or quantum entanglement. By long range we mean length scale of the order of inverse temperature or bigger.  So if we integrate out UV degrees of freedom and come down to the thermal scale, \textit{the effective quantum state in the IR will be separable or unentangled}. This state is \textit{classical} from the point of view of quantum information theory. Entanglement or Von-Neumann entropy in this state reduces to thermal entropy and is a \textit{measure of classical or thermal correlations}. Along the RG-flow the effective temperature also grows because temperature is a relevant parameter. This effective temperature can be thought of as the local temperature measured by a stationary observer in the bulk at a certain radial distance from the boundary. The temperature diverges as we move closer to the horizon. So \textit{these high temperature IR degrees of freedom of the field theory can be thought of as "located" in the near horizon region in the bulk}. One way to visulaize this is to compute the (renormalized) entanglement entropy in the IR. In the large-N limit one can use Ryu-Takayanagi prescription \cite{Ryu:2006bv} to perform this computation. One has to take a subsystem of size much bigger than the thermal scale. For such a subsystem the dominant contribution to the entanglement entropy comes from the portion of the minimal surface in the bulk which touches the horizon. This is clear from the fact that for subsystem size much bigger than the thermal scale entanglement entropy crosses over to the thermal entropy. The important point is that the minimal surface for the stationary thermal state does not cross the horizon. An interesting description of this RG-flow appears in \cite{Swingle:2009bg} from the point of view of entanglement renormalization. 

\section{Information theoretic \textit{distinction} between horizon and singularity}

\begin{figure}[htbp]
\begin{center}
  \includegraphics[width=13cm]{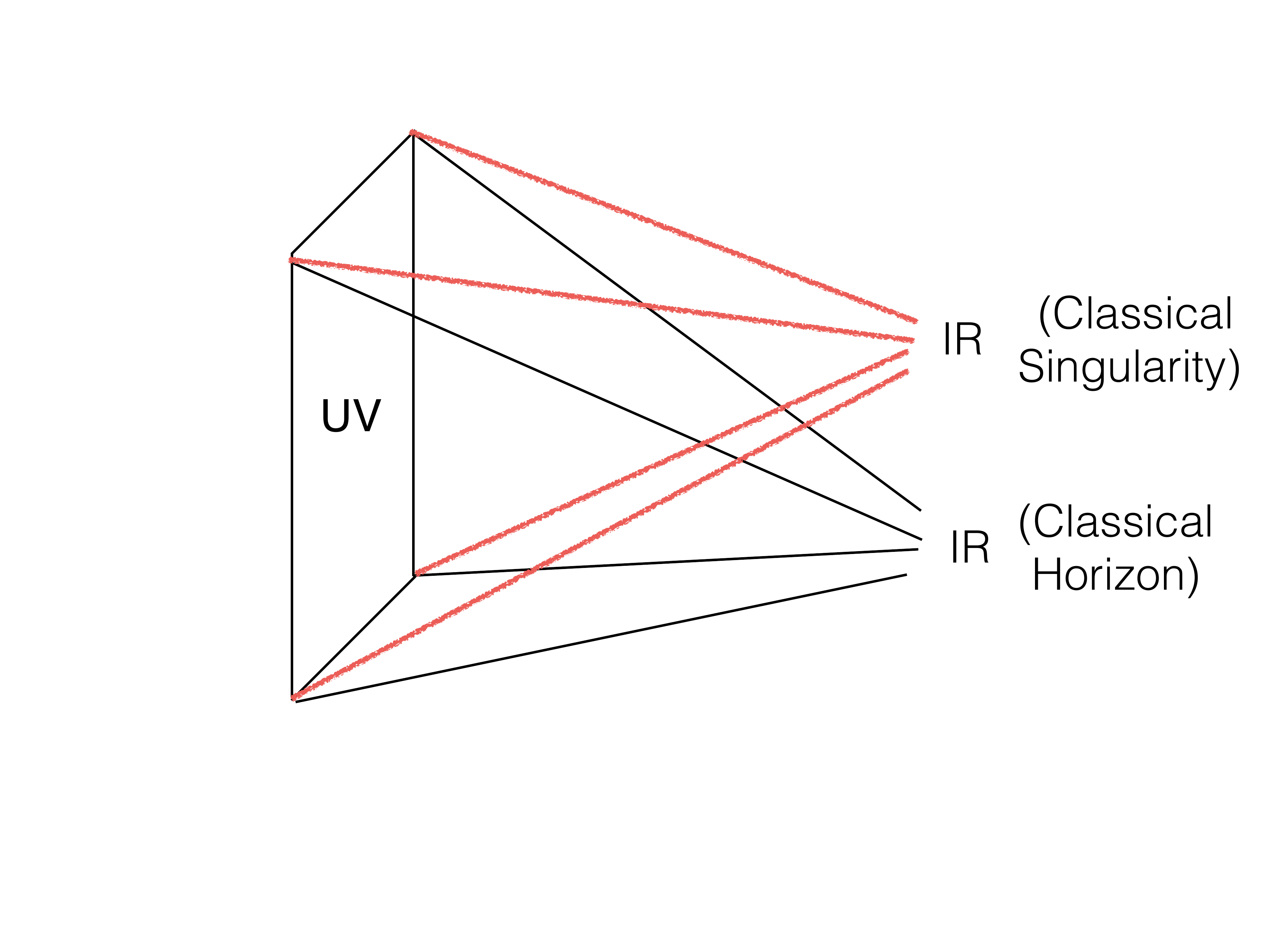}
\end{center}
\vspace{-8em}
\caption{This is a cartoon of the Schwarzschild black brane geometry from the point of view of holographic-RG when we treat the classical and quantum correlations as "observables". There is a single UV boundary but two \textit{distinct} IR regions in the interior corresponding to RG-flow in asymptotic (black) and infalling (red) frames. So "locality" in the "radial direction" or "energy-scale" breaks down in a very specific manner. We will argue that this is essentially a "classical".}
\end{figure}

Fig-2 summarizes our previous discussion. RG flow in a thermal state has two different bulk descriptions which are apparently contradictory. In one description where we consider the flow of classical correlations (e.g, entanglement or Von-Neumann entropy) the IR of the field theory can be identified with the near-horizon region whereas in a different description where we consider the flow of quantum correlations (e.g, some entanglement measure for mixed states like negativity) the IR can be identified with the region near the curvature singularity. It is natural to associate the first description with the asymptotic observer and the second one with the infalling observer.

\textit{The upshot of this whole discussion is that the very long wave-length degrees of freedom in a thermal state carry information about the near-horizon as well as the near-singularity regions. In other words, near-horizon and near-singularity regions can be thought of as manifestations of two different aspects (e.g, classical and quantum correlations, respectively) of the \textit{same} IR degrees of freedom in the field theory}. \textit{Classical and quantum correlation} is a pair of information theoretic quantities which can \textit{distinguish} between near-horizon and near-singularity regions in the thermal CFT  but there can be many more such pairs. But, although classical and quantum correlations can be quantified, there are no standard quantum mechanical observables which can measure them.  

\section{Towards Resolution of The Classical Singularity}  

Since string theory is a consistent theory of quantum gravity, it should be able to make sense of classical black hole singularity. This does not include, for example, the singularity of \textit{negative mass} Schwarzschild solution \cite{Horowitz:1995ta} but curvature singularity \textit{hidden behind a horizon} should be resolved in string theory. What we mean by resolved is that if we ask a \textit{physical question} about the singularity in string theory then we should get a meaningful finite answer \cite{Liu:2002yd,Liu:2002ft}. In general the set of such "physical questions" is difficult to determine. One of the potential difficulties is that a question which is physical within the framework of effective field theory may not remain so when embedded in non-perturbative string theory. We will now give some heuristic arguments which suggest that this is indeed the case, at least for Schwarzschild black brane (hole) in AdS.  

AdS-CFT duality tell us that the observables of quantum gravity in asymptotically AdS space-times are the correlation functions of the boundary CFT (or string scattering amplitudes in the bulk). \textit{Therefore the answer to a physical question in bulk quantum gravity can be obtained by computing some set of correlation functions in the CFT. So if the CFT correlation functions do not contain answer to some bulk question then we will say that the question is not physical}. 

\begin{figure}[htbp]
\begin{center}
  \includegraphics[width=13cm]{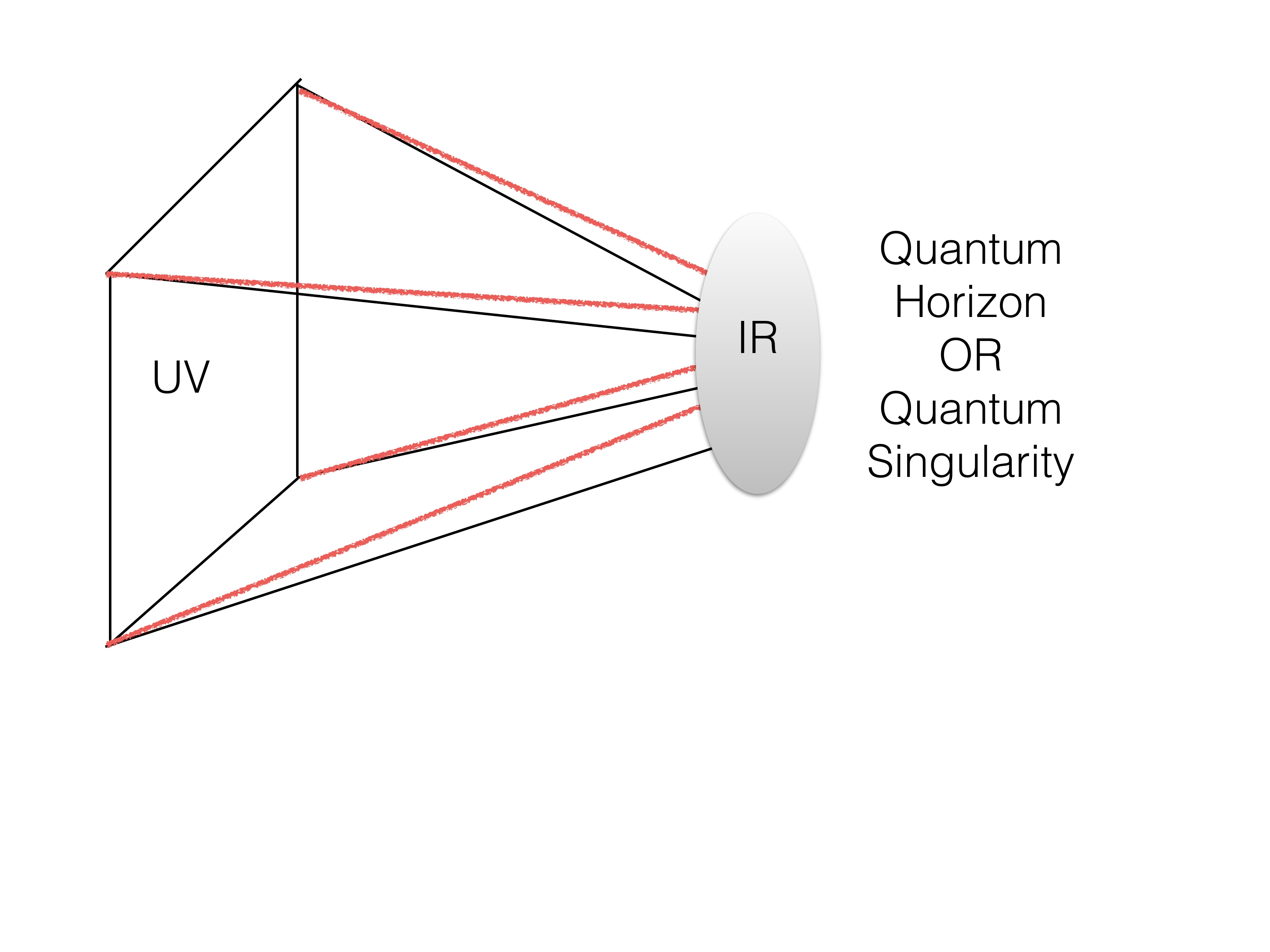}
\end{center}
\vspace{-8em}
\caption{This is a cartoon of the Schwarzschild black brane geometry from the point of view of holographic-RG when we use the \textit{exact observables} of the theory which are the \textit{correlation functions of the CFT}. Due to \textit{non-separability of correlators}, two distinct IR regions (classical horizon and classical singularity) in Fig-2 have been replaced by a \textit{single} IR region which may either be called quantum horizon (in the asymptotic frame) or quantum singularity (in the infalling frame). This suggests that classical singularity \textit{by itself} is \textit{not} a physical problem.}
\end{figure}

We have seen that in the thermal CFT the information about the near-singularity region is encoded in the very long wave-length (IR) degrees of freedom. But it is \textit{also} true that the \textit{same} long wave-length degrees of freedom carry information about the near-horizon region. So let us consider a thermal correlator in the CFT and focus on a kinematic regime where the correlator is able to probe the very long-wavelength degrees of freedom in the thermal state. To extract information about the near-singularity region from the correlator one has to \textit{subtract} the contribution of the near-horizon region. In other words, if we want to extract information about the near-singularity region we need to have a \textit{unique} decomposition of the value of the correlator into two parts, one of which has information about the near-horizon region and the other one about the near-singularity region. \textit{But such a unique decomposition of correlation functions is not meaningful or rather not observable.} The simplest reason being that CFT (or any QFT) gives us a prescription for computing the value of a correlation function only. In this particular context, we have to further decompose the value into two parts based on some criterion determined by say the properties of (thermal) classical ($\rightarrow$ horizon) and quantum ($\rightarrow$ singularity) correlations (entanglement) of the IR degrees of freedom. But, due to the principle of linear superposition of quantum states, there are no standard quantum mechanical observables that can \textit{measure} such correlations \cite{Maldacena:2013xja}. \footnote{See also \cite{Jafferis:2017tiu} for an interesting discussion on some related issues.} This suggests that \textit{a field theory observer cannot, in principle, disentangle the information of the near-singularity region from the information of the near-horizon region, that is contained in the CFT correlation functions}. \footnote{This is somewhat similar to the following situation. Suppose we are given the cross-section for $2\rightarrow 2$ scattering of electrons in QED. Now can we determine that how much of the cross-section is due to the particle nature and how much is due to the wave nature of the electrons ? The answer is obviously no as long as we use the mathematical framework of quantum mechanics to calculate the cross-section.} As a result of this, from the CFT point of view, both \textit{the near-horizon and the near-singularity regions completely lose their separate identities}. This can also be stated as : from the CFT point of view near-horizon and near-singularity regions are \textit{non-separable}. In the rest of the paper we will refer to this as "\textbf{non-separability of CFT correlation functions}". Therefore in the non-perturbative CFT description of the black hole, the concept of the \textit{near-singularity} region, by itself, completely \textit{disappears} from the theory and the same is true for the near-horizon region. Instead, they are replaced by a \textit{single concept} (Fig-3) which may be called either the \textit{quantum} horizon \textit{or} the \textit{quantum} singularity depending on whether we choose the \textit{asymptotic} or the \textit{infalling} frame for the bulk interpretation of the CFT correlation functions. This should be contrasted with the description of the black hole in the framework of local effective field theory where the near-horizon and the near-singularity regions are clearly distinct space-time regions with very different physical properties.

We would like to emphasize that this does not mean that the horizon is replaced by singularity or vice-versa. Rather this should be interpreted as a very specific type of "non-locality" in the bulk which "\textit{relates} the observations of the asymptotic and the infalling observers". This is actually a statement about the nature of observables in AdS quantum gravity (or perhaps in any holographic \cite{Maldacena:1997re,Witten:1998qj,Gubser:1998bc,Banks:1996vh,Susskind:1994vu,tHooft:1993dmi, Bousso:1999xy} or asymptotic description) and so it concerns only the \textit{kinematics} of the theory. Also the fact that the quantum gravity in the bulk is dual to a standard \textit{linear} quantum mechanical system is crucial for this \textit{relation} to exist. 

Although we are interpreting the \textit{non-separability of CFT correlators} as a kind of "non-locality" in the bulk, this is actually a \textit{misnomer}. In particular, we would like to emphasize that this is \textit{not} the type of non-locality which can be described as a "non-local theory living on a background space-time and the result of non-locality is signal propagation outside the light-cone." We hope to clarify this more in later sections.

Let us now discuss the implications for the classical singularity behind the horizon which is the main focus of this paper. 

For the time being let us think about the black hole from a global point of view. This may also be called the "nice-slice point of view". This description is refereed to an imaginary observer who can see all of the Penrose diagram. Now from effective field theory point of view singularity is a UV problem. Since curvature in the near-singularity region can be of the order of the Planck scale, large stringy and quantum effects are expected in this region. On the contrary, the near-horizon region of a sufficiently large black hole is almost flat space and so the physics in this region is expected to be well-described by the standard low energy theory. Now we have already described in the previous sections that from the holographic RG point of view singularity can be thought of as a \textit{trivial IR fixed-point of a gapped system}. So a natural guess could be that the new stringy degrees of freedom, \textit{which live near the singularity}, "smooth it out" to a kind of Planck scale fuzz. These new degrees of freedom are precisely those missed by pure gravity. 

The whole point of the above discussion is to emphasize the important point that in the framework of effective field theory the usual \textit{questions} about the \textit{classical singularity} and their expected \textit{answers} are all \textit{local} in nature. Therefore one can use these local bulk \textit{questions} and \textit{answers} to \textit{single out} the near-singularity region in the black hole geometry. But we can easily convince ourselves that this is in \textit{sharp contradiction} with the \textit{non-separability} of the CFT corrleation functions. Due to the same reason, \textit{local} bulk statements like "space-time curvature blows up \textit{at the singularity}" or "quantum fluctuation of the metric becomes large \textit{near the singularity}" are also \textit{unphysical} from the CFT point of view. It is important to note that "unphysical" does \textit{not} mean \textit{wrong}. What it means is that within the framework of holography it is \textit{not} possible to decide, \textit{in principle}, whether the above statements are true or false. From the bulk point of view we would like to say that, \textit{classical black hole singularity, by itself, is not a physical problem in quantum gravity.} Very roughly speaking \textit{the "non-locality" of the holographic description in the presence of a black hole does not allow us to zoom in near the singularity and as a result the classical singularity, which is essentially local in nature, is completely dissolved}. We have used the term "dissolved", rather than "resolved", to emphasize the fact that the classical singularity does not appear to be removed or smoothed out by any dynamical mechanism. But what is happening is that the \textit{singularity as a local space-time concept} has completely disappeared from the theory due to "mixing" with the horizon. The reason for this seems to be the linearity of the dual description and the fact that quantum gravity is holographic in nature.

Now from the bulk point of view, the reason that such local or EFT questions about the classical singularity cannot be posed in quantum gravity, seems to be related to a \textit{fundamental limitation on the type of local measurements that an infalling observer can perform}. This limitation has nothing to do with quantum mechanical uncertainty in the bulk, but appears to be closely related to the \textit{holographic bound on the number of degrees of freedom in quantum gravity.} Non-separability of CFT correlation functions is a reflection of this limitation. The standard picture of local space-time physics can be trusted in the regime where this limitation can be ignored in the same way that classical description is trustworthy in the regime where quantum mechanical uncertainty can be ignored. Although we do not know what this limitation on bulk measurements is but the duality is powerful enough to let us guess some of its consequences and do precise calculations. 

\section{Diagnostics for the "Classical-Singularity"}

From the previous discussions it is clear that for an unambiguous space-time interpretation of CFT correlation functions we have to refer to a particular bulk observer because the global description is not consistent with the non-separability of correlators. This is an important point. But, as we will discuss, even with the choice of a particular observer \textit{space-time interpretation is only approximate} if our arguments are correct. In other words, \textit{exact observables do not have exact space-time interpretation}. The reason is that there are always some effects, that an observer can detect, which have no conventional space-time interpretation. So let us discuss the asymptotic observer. 


Although our arguments in the previous sections involving (holographic)RG-flow is very well-suited for black branes, we will assume that the non-separability of CFT correlation functions is also true for an AdS-Schwarschild black hole with compact horizon. Another way to think about this will be that the CFT lives on a sphere of radius $R$ and the temperature $T$ is $>> R^{-1}$. Now we will give a heuristic argument to suggest that \textit{at late time the thermal correlation functions contain contributions that can be thought of as "coming from the region near the classical singularity".} In particular we will consider the thermal two-point function which is also a diagnostic for information loss in AdS-CFT \cite{Maldacena:2001kr}. 

Let us consider the thermal two-point function $<O(t)O(0)>_{th}$ where $O$ is some operator in the CFT.  The operator $O(0)$ creates an excitation in the bulk at Schwarschild time $t=0$ which then falls towards the black hole. With respect to the asymptotic observer the excitation does not cross the horizon in finite Schwarschild time, but after a sufficiently long time we can safely assume that the excitation starts probing the near-horizon region of the black hole. Now according to our previous discussions, if we use the exact observables of the theory which are the CFT correlation functions or string scattering amplitude in the bulk, then the near-horizon and the near-singularity regions are non-separable. Now \textit{owing to this non-separability of correlators}, an excitation which is in the near-horizon region \textit{can also be thought of} as "probing" the region near the singularity. So the late time limit of a thermal correlation function, in particular the two-point function, has contributions "coming from the region near the classical singularity". 

Although the late time value of the two-point function has contributions that can be thought of as coming from the "near-singularity region", the asymptotic observer will naturally interpret them as some effects which originate near the horizon. The point here is that \textit{this "horizon", at late time, can no longer be completely described by the standard low energy effective field theory in the asymptotic frame because of its "mixing" with the singularity}. So the \textit{horizon} that the asymptotic observer observes (at late times) is \textit{not} the "horizon of the effective field theory" but may be called the "quantum horizon" as we have discussed in the last section. From the point of view of global effective field theory description of black hole this may be called "IR/UV mixing" where IR refers to the classical near-horizon region and UV refers to the classical near-singularity region where Planck scale effects are supposed to be dominant. This mixing is visible to a low energy asymptotic observer. Now what is the order of magnitude of such effects. This can be estimated by looking at the \textit{deviation} of the CFT correlation function from the bulk effective field theory prediction at late time and it is natural to associate some of the O$(e^{-S_{BH}} \sim e^{-N^2})$- effects \cite{Maldacena:2001kr,Cotler:2016fpe,Fitzpatrick:2016ive,Fitzpatrick:2016mjq, Anous:2016kss, Barbon:2003aq} with the "classical singularity behind the horizon". So in the large-N limit this can be a very small effect. But the important point is that this effect from "behind the horizon" is required for the consistency of the holographic description of black hole from the point of view of the asymptotic observers. 

The reader may be worried by an apparent violation of causality (and locality) in the above description. But let us emphasize that the \textit{"information" about the singularity is not physically or dynamically transferred from the near-singularity region to the near-horizon region because physical transfer implies that we are able to distinguish between the near-horizon and the near-singularity regions. But our previous discussions suggest that this distinction is unphysical due to the non-separability of CFT correlators.} So we should not describe this as causality-violation. This is different from the situation in quantum field theory (QFT) where non-locality generically leads to causality violation. In fact strictly speaking these effects may not have any space-time interpretation in the sense of \textit{canonical QG}.

Before we conclude we would like to emphasize that whatever we have said so far is strictly meant for a black hole formed from collapse.  For a two-sided black hole the story seems to be quite different \cite{Kraus:2002iv,Fidkowski:2003nf,Festuccia:2005pi}. We do not completely understand the reason behind this difference. But, it is important to have some understanding of this. A first step in this direction will be to generalize our information-theoretic approach to the two-sided case perhaps using the idea of computational complexity \cite{Brown:2015bva} and various bulk reconstruction methods \cite{Hamilton:2006az, Papadodimas:2012aq,Almheiri:2014lwa,Dong:2016eik,Mintun:2015qda} that have been studied in the literature.

\section{Fate of The Singularity (?)}

If our arguments are correct then the "classical singularity" is \textit{dissolved} due to the "non-locality" of the holographic description. 

Our arguments further suggest that in the large-N limit quantum gravity \textit{effectively} makes the black hole singularity \textit{"slightly naked".} This, in a sense, is consistent with the Cosmic censorship. The asymptotic observer does not see any "violent" Planck scale effect. Instead, due to the absence of local degrees of freedom or the holographic bound on the number of degrees of freedom in quantum gravity, the "Planck scale effects in the near-singularity region" are turned into "soft ($\sim e^{-N^2}$) IR effects in the almost flat near-horizon region". This is then described by the asymptotic observer as \textit{deviation} from the bulk low-energy effective field theory prediction at late time. And moreover, \textit{the completeness of the CFT description} tells us that such deviations carry \textit{complete information} about the "Planck scale effects near the singularity". \textit{The surprising fact is that this should now be accessible to low energy observers, at least in principle}. Calculation of such effects requires knowledge of the \textit{non-perturbative theory} which in this case is the dual CFT. 

Another way of saying this will be that the outside description is a \textit{complete} description \textit{although} there are some late-time effects which, in the \textit{conventional} global space-time description, can be thought of as coming from the "region near the singularity". But, as we have already argued, this, in a sense, is "non-local" but \textit{not causality-violating}. Hopefully the recent progress in understanding the late-time behaviour of thermal correlators \cite{Cotler:2016fpe,Fitzpatrick:2016ive,Fitzpatrick:2016mjq, Anous:2016kss} will shed more light on the nature of such effects.





\section{Some comments on black hole complementarity}

The reader may have noticed that our observations so far are strikingly similar to the Black Hole Complementarity idea \cite{Stephens:1993an,Susskind:1993if,Susskind:1993mu} which says that the space-time location of the information depends on the choice of the observer. In the asymptotic frame the infalling matter ends up in the near-horizon region whereas in the infalling frame the same matter smoothly passes through the horizon and finally hits the singularity. In spite of this, no single observer should be able to know about both the end-points. This leaves open the possibility that an imaginary global observer who can see all of the Penrose diagram can apparently see "duplication" of the information. Since no such observer exists, the question is can the theory, which we use to calculate say the unitary S-matrix for an evaporating black hole, allow \textit{two distinct} fates of the infalling matter? If we now apply the arguments of this paper then the answer is clearly \textit{no}. We can easily convince ourselves that non-separability of correlation functions does \textit{not} allow the CFT to \textit{distinguish between these two fates}. This is a schematic argument but we hope to have conveyed the sense in which "duplication" does not happen in QG and it is very closely related to the fact that the bulk is dual to a standard \textit{linear} quantum mechanical theory. It will be very interesting to connect the picture that we have tried to produce with the one given in \cite{Maldacena:2017axo}.  

We would also like to know if the arguments in this paper have anything to say about the firewall paradox \cite{Almheiri:2012rt,Mathur:2009hf}. We hope to have something to say about this in future.


\section{Acknowledgement}

It is a pleasure to thank Dionysios Anninos and Rajesh Gopakumar for helpful discussions and Juan Maldacena for pointing out the relation between state-dependence and non-linearity. I would also like to thank Pankaj Agrawal, Pallab Basu, Chetan Krishnan, Arnab kundu, Gautam Mandal, Shiraz Minwalla, Sudipta Mukherjee, Djordje Radicevic, Shubho Roy, Ashoke Sen, Amitabh Virmani and Aron Wall for discussions on related matters.


\begin{thebibliography}{99}
  
  \bibitem{Maldacena:1997re} 
  J.~M.~Maldacena,
  ``The Large N limit of superconformal field theories and supergravity,''
  Int.\ J.\ Theor.\ Phys.\  {\bf 38}, 1113 (1999)
  [Adv.\ Theor.\ Math.\ Phys.\  {\bf 2}, 231 (1998)]
  doi:10.1023/A:1026654312961
  [hep-th/9711200].
   
  \bibitem{Witten:1998qj} 
  E.~Witten,
  ``Anti-de Sitter space and holography,''
  Adv.\ Theor.\ Math.\ Phys.\  {\bf 2}, 253 (1998)
  [hep-th/9802150].
  
  \bibitem{Gubser:1998bc} 
  S.~S.~Gubser, I.~R.~Klebanov and A.~M.~Polyakov,
  ``Gauge theory correlators from noncritical string theory,''
  Phys.\ Lett.\ B {\bf 428}, 105 (1998)
  doi:10.1016/S0370-2693(98)00377-3
  [hep-th/9802109].
  
  \bibitem{Banks:1996vh} 
  T.~Banks, W.~Fischler, S.~H.~Shenker and L.~Susskind,
  ``M theory as a matrix model: A Conjecture,''
  Phys.\ Rev.\ D {\bf 55}, 5112 (1997)
  doi:10.1103/PhysRevD.55.5112
  [hep-th/9610043].
  
  \bibitem{Susskind:1994vu} 
  L.~Susskind,
  ``The World as a hologram,''
  J.\ Math.\ Phys.\  {\bf 36}, 6377 (1995)
  doi:10.1063/1.531249
  [hep-th/9409089]. 
  
 \bibitem{tHooft:1993dmi} 
  G.~'t Hooft,
  ``Dimensional reduction in quantum gravity,''
  Salamfest 1993:0284-296
  [gr-qc/9310026].
  
  \bibitem{Bousso:1999xy} 
  R.~Bousso,
  ``A Covariant entropy conjecture,''
  JHEP {\bf 9907}, 004 (1999)
  doi:10.1088/1126-6708/1999/07/004
  [hep-th/9905177].
  
  R.~Bousso,
  ``Holography in general space-times,''
  JHEP {\bf 9906}, 028 (1999)
  doi:10.1088/1126-6708/1999/06/028
  [hep-th/9906022].
  
  \bibitem{Susskind:1998dq} 
  L.~Susskind and E.~Witten,
  ``The Holographic bound in anti-de Sitter space,''
  hep-th/9805114.
    
  \bibitem{Maldacena:2001kr} 
  J.~M.~Maldacena,
  ``Eternal black holes in anti-de Sitter,''
  JHEP {\bf 0304}, 021 (2003)
  doi:10.1088/1126-6708/2003/04/021
  [hep-th/0106112].

    \bibitem{Cotler:2016fpe} 
  J.~S.~Cotler {\it et al.},
  ``Black Holes and Random Matrices,''
  arXiv:1611.04650 [hep-th]. 
  
  \bibitem{Fitzpatrick:2016ive} 
  A.~L.~Fitzpatrick, J.~Kaplan, D.~Li and J.~Wang,
  ``On information loss in AdS$_{3}$/CFT$_{2}$,''
  JHEP {\bf 1605}, 109 (2016)
  doi:10.1007/JHEP05(2016)109
  [arXiv:1603.08925 [hep-th]].
  
  \bibitem{Fitzpatrick:2016mjq} 
  A.~Liam Fitzpatrick and J.~Kaplan,
  ``On the Late-Time Behavior of Virasoro Blocks and a Classification of Semiclassical Saddles,''
  arXiv:1609.07153 [hep-th].
  
  \bibitem{Anous:2016kss} 
  T.~Anous, T.~Hartman, A.~Rovai and J.~Sonner,
  ``Black Hole Collapse in the 1/c Expansion,''
  JHEP {\bf 1607}, 123 (2016)
  doi:10.1007/JHEP07(2016)123
  [arXiv:1603.04856 [hep-th]]. 
  
   \bibitem{Barbon:2003aq} 
  J.~L.~F.~Barbon and E.~Rabinovici,
  ``Very long time scales and black hole thermal equilibrium,''
  JHEP {\bf 0311}, 047 (2003)
  doi:10.1088/1126-6708/2003/11/047
  [hep-th/0308063].
  
   \bibitem{Kraus:2002iv} 
  P.~Kraus, H.~Ooguri and S.~Shenker,
  ``Inside the horizon with AdS / CFT,''
  Phys.\ Rev.\ D {\bf 67}, 124022 (2003)
  doi:10.1103/PhysRevD.67.124022
  [hep-th/0212277].
  
  \bibitem{Fidkowski:2003nf} 
  L.~Fidkowski, V.~Hubeny, M.~Kleban and S.~Shenker,
  ``The Black hole singularity in AdS / CFT,''
  JHEP {\bf 0402}, 014 (2004)
  doi:10.1088/1126-6708/2004/02/014
  [hep-th/0306170].  
  
  \bibitem{Festuccia:2005pi} 
  G.~Festuccia and H.~Liu,
  ``Excursions beyond the horizon: Black hole singularities in Yang-Mills theories. I.,''
  JHEP {\bf 0604}, 044 (2006)
  doi:10.1088/1126-6708/2006/04/044
  [hep-th/0506202].
  
   \bibitem{Liu:2002yd} 
  H.~Liu, G.~W.~Moore and N.~Seiberg,
  ``The Challenging cosmic singularity,''
  gr-qc/0301001.
  
  \bibitem{Liu:2002ft} 
  H.~Liu, G.~W.~Moore and N.~Seiberg,
  ``Strings in a time dependent orbifold,''
  JHEP {\bf 0206}, 045 (2002)
  doi:10.1088/1126-6708/2002/06/045
  [hep-th/0204168].
  
  \bibitem{Horowitz:1995ta} 
  G.~T.~Horowitz and R.~C.~Myers,
  ``The value of singularities,''
  Gen.\ Rel.\ Grav.\  {\bf 27}, 915 (1995)
  doi:10.1007/BF02113073
  [gr-qc/9503062].

 \bibitem{Banerjee:2015coc} 
  S.~Banerjee and P.~Paul,
  ``Black Hole Singularity, Generalized (Holographic) $c$-Theorem and Entanglement Negativity,''
  JHEP {\bf 1702}, 043 (2017)
  doi:10.1007/JHEP02(2017)043
  [arXiv:1512.02232 [hep-th]].
  
   \bibitem{Akhmedov:1998vf} 
  E.~T.~Akhmedov,
  ``A Remark on the AdS / CFT correspondence and the renormalization group flow,''
  Phys.\ Lett.\ B {\bf 442}, 152 (1998)
  [hep-th/9806217].

  D.~Z.~Freedman, S.~S.~Gubser, K.~Pilch and N.~P.~Warner,
  ``Renormalization group flows from holography supersymmetry and a c theorem,''
  Adv.\ Theor.\ Math.\ Phys.\  {\bf 3}, 363 (1999)
  [hep-th/9904017].

  L.~Girardello, M.~Petrini, M.~Porrati and A.~Zaffaroni,
  ``Novel local CFT and exact results on perturbations of N=4 superYang Mills from AdS dynamics,'' 
 JHEP {\bf 9812}, 022 (1998)  [hep-th/9810126].
 
  J.~de Boer, E.~P.~Verlinde and H.~L.~Verlinde,
  ``On the holographic renormalization group,''
  JHEP {\bf 0008}, 003 (2000)
  [hep-th/9912012].
  
R.~C.~Myers and A.~Sinha,
  ``Holographic c-theorems in arbitrary dimensions,''
  JHEP {\bf 1101}, 125 (2011)
  [arXiv:1011.5819 [hep-th]].

  
 \bibitem{sb} 
   S.~Banerjee,
  ``RG Flow and Thermodynamics of Causal Horizons in AdS,''
  JHEP {\bf 1510}, 098 (2015)
  [arXiv:1508.01343 [hep-th]].
  
  \bibitem{Banerjee:2015uaa} 
  S.~Banerjee and A.~Bhattacharyya,
  ``RG Flow and Thermodynamics of Causal Horizons in Higher-Derivative AdS Gravity,''
  arXiv:1509.08475 [hep-th].
  
  \bibitem{Hartman:2013qma} 
  T.~Hartman and J.~Maldacena,
  ``Time Evolution of Entanglement Entropy from Black Hole Interiors,''
  JHEP {\bf 1305}, 014 (2013)
  doi:10.1007/JHEP05(2013)014
  [arXiv:1303.1080 [hep-th]]
    
 \bibitem{Susskind:1993if} 
  L.~Susskind, L.~Thorlacius and J.~Uglum,
  ``The Stretched horizon and black hole complementarity,''
  Phys.\ Rev.\ D {\bf 48}, 3743 (1993)
  doi:10.1103/PhysRevD.48.3743
  [hep-th/9306069].
  
  \bibitem{Susskind:1993mu} 
  L.~Susskind and L.~Thorlacius,
  ``Gedanken experiments involving black holes,''
  Phys.\ Rev.\ D {\bf 49}, 966 (1994)
  doi:10.1103/PhysRevD.49.966
  [hep-th/9308100].

 
   \bibitem{Stephens:1993an} 
  C.~R.~Stephens, G.~'t Hooft and B.~F.~Whiting,
  ``Black hole evaporation without information loss,''
  Class.\ Quant.\ Grav.\  {\bf 11}, 621 (1994)
  doi:10.1088/0264-9381/11/3/014
  [gr-qc/9310006].
          
  \bibitem{Kiem:1995iy} 
  Y.~Kiem, H.~L.~Verlinde and E.~P.~Verlinde,
  ``Black hole horizons and complementarity,''
  Phys.\ Rev.\ D {\bf 52}, 7053 (1995)
  doi:10.1103/PhysRevD.52.7053
  [hep-th/9502074].
    
\bibitem{Susskind:1993aa} 
  L.~Susskind,
  ``Strings, black holes and Lorentz contraction,''
  Phys.\ Rev.\ D {\bf 49}, 6606 (1994)
  doi:10.1103/PhysRevD.49.6606
  [hep-th/9308139].
  
  L.~Susskind,
  ``String theory and the principles of black hole complementarity,''
  Phys.\ Rev.\ Lett.\  {\bf 71}, 2367 (1993)
  doi:10.1103/PhysRevLett.71.2367
  [hep-th/9307168].
  
  D.~A.~Lowe, J.~Polchinski, L.~Susskind, L.~Thorlacius and J.~Uglum,
  ``Black hole complementarity versus locality,''
  Phys.\ Rev.\ D {\bf 52}, 6997 (1995)
  doi:10.1103/PhysRevD.52.6997
  [hep-th/9506138].
  
  \bibitem{Almheiri:2012rt} 
  A.~Almheiri, D.~Marolf, J.~Polchinski and J.~Sully,
  ``Black Holes: Complementarity or Firewalls?,''
  JHEP {\bf 1302}, 062 (2013)
  doi:10.1007/JHEP02(2013)062
  [arXiv:1207.3123 [hep-th]].
  
  A.~Almheiri, D.~Marolf, J.~Polchinski, D.~Stanford and J.~Sully,
  ``An Apologia for Firewalls,''
  JHEP {\bf 1309}, 018 (2013)
  doi:10.1007/JHEP09(2013)018
  [arXiv:1304.6483 [hep-th]].
  
  \bibitem{Mathur:2009hf} 
  S.~D.~Mathur,
  ``The Information paradox: A Pedagogical introduction,''
  Class.\ Quant.\ Grav.\  {\bf 26}, 224001 (2009)
  doi:10.1088/0264-9381/26/22/224001
  [arXiv:0909.1038 [hep-th]].
      
  \bibitem{Maldacena:2013xja} 
  J.~Maldacena and L.~Susskind,
  ``Cool horizons for entangled black holes,''
  Fortsch.\ Phys.\  {\bf 61}, 781 (2013)
  doi:10.1002/prop.201300020
  [arXiv:1306.0533 [hep-th]].
  
  \bibitem{Maldacena:2017axo} 
  J.~Maldacena, D.~Stanford and Z.~Yang,
  ``Diving into traversable wormholes,''
  Fortsch.\ Phys.\  {\bf 65}, no. 5, 1700034 (2017)
  doi:10.1002/prop.201700034
  [arXiv:1704.05333 [hep-th]].
  
  \bibitem{Brown:2015bva} 
  A.~R.~Brown, D.~A.~Roberts, L.~Susskind, B.~Swingle and Y.~Zhao,
  ``Holographic Complexity Equals Bulk Action?,''
  Phys.\ Rev.\ Lett.\  {\bf 116}, no. 19, 191301 (2016)
  doi:10.1103/PhysRevLett.116.191301
  [arXiv:1509.07876 [hep-th]].
  
  A.~R.~Brown, D.~A.~Roberts, L.~Susskind, B.~Swingle and Y.~Zhao,
  ``Complexity, action, and black holes,''
  Phys.\ Rev.\ D {\bf 93}, no. 8, 086006 (2016)
  doi:10.1103/PhysRevD.93.086006
  [arXiv:1512.04993 [hep-th]].
  
  \bibitem{Hamilton:2006az} 
  A.~Hamilton, D.~N.~Kabat, G.~Lifschytz and D.~A.~Lowe,
  ``Holographic representation of local bulk operators,''
  Phys.\ Rev.\ D {\bf 74}, 066009 (2006)
  doi:10.1103/PhysRevD.74.066009
  [hep-th/0606141].
  
  A.~Hamilton, D.~N.~Kabat, G.~Lifschytz and D.~A.~Lowe,
  ``Local bulk operators in AdS/CFT: A Holographic description of the black hole interior,''
  Phys.\ Rev.\ D {\bf 75}, 106001 (2007)
  Erratum: [Phys.\ Rev.\ D {\bf 75}, 129902 (2007)]
  doi:10.1103/PhysRevD.75.106001, 10.1103/PhysRevD.75.129902
  [hep-th/0612053].
    
  \bibitem{Papadodimas:2012aq} 
  K.~Papadodimas and S.~Raju,
  ``An Infalling Observer in AdS/CFT,''
  JHEP {\bf 1310}, 212 (2013)
  doi:10.1007/JHEP10(2013)212
  [arXiv:1211.6767 [hep-th]].
  
  K.~Papadodimas and S.~Raju,
  ``State-Dependent Bulk-Boundary Maps and Black Hole Complementarity,''
  Phys.\ Rev.\ D {\bf 89}, no. 8, 086010 (2014)
  doi:10.1103/PhysRevD.89.086010
  [arXiv:1310.6335 [hep-th]].
  
  K.~Papadodimas and S.~Raju,
  ``Local Operators in the Eternal Black Hole,''
  Phys.\ Rev.\ Lett.\  {\bf 115}, no. 21, 211601 (2015)
  doi:10.1103/PhysRevLett.115.211601
  [arXiv:1502.06692 [hep-th]].
  
  \bibitem{Almheiri:2014lwa} 
  A.~Almheiri, X.~Dong and D.~Harlow,
  ``Bulk Locality and Quantum Error Correction in AdS/CFT,''
  JHEP {\bf 1504}, 163 (2015)
  doi:10.1007/JHEP04(2015)163
  [arXiv:1411.7041 [hep-th]].
  
  F.~Pastawski, B.~Yoshida, D.~Harlow and J.~Preskill,
  ``Holographic quantum error-correcting codes: Toy models for the bulk/boundary correspondence,''
  JHEP {\bf 1506}, 149 (2015)
  doi:10.1007/JHEP06(2015)149
  [arXiv:1503.06237 [hep-th]].
  
  \bibitem{Dong:2016eik} 
  X.~Dong, D.~Harlow and A.~C.~Wall,
  ``Reconstruction of Bulk Operators within the Entanglement Wedge in Gauge-Gravity Duality,''
  Phys.\ Rev.\ Lett.\  {\bf 117}, no. 2, 021601 (2016)
  doi:10.1103/PhysRevLett.117.021601
  [arXiv:1601.05416 [hep-th]].
  
  \bibitem{Mintun:2015qda} 
  E.~Mintun, J.~Polchinski and V.~Rosenhaus,
  ``Bulk-Boundary Duality, Gauge Invariance, and Quantum Error Corrections,''
  Phys.\ Rev.\ Lett.\  {\bf 115}, no. 15, 151601 (2015)
  doi:10.1103/PhysRevLett.115.151601
  [arXiv:1501.06577 [hep-th]].
  
  \bibitem{Swingle:2009bg} 
  B.~Swingle,
  ``Entanglement Renormalization and Holography,''
  Phys.\ Rev.\ D {\bf 86}, 065007 (2012)
  doi:10.1103/PhysRevD.86.065007
  [arXiv:0905.1317 [cond-mat.str-el]]. 

  
  \bibitem{Jafferis:2017tiu} 
  D.~L.~Jafferis,
  ``Bulk reconstruction and the Hartle-Hawking wavefunction,''
  arXiv:1703.01519 [hep-th].
  
  \bibitem{Zam}
  A.~B.~Zamolodchikov,
  ``Irreversibility of the Flux of the Renormalization Group in a 2D Field Theory,''
  JETP Lett.\  {\bf 43} (1986) 730
   [Pisma Zh.\ Eksp.\ Teor.\ Fiz.\  {\bf 43} (1986) 565].
   
  J.~L.~Cardy,
``Is There a c Theorem in Four-Dimensions?,''
  Phys.\ Lett.\ B {\bf 215} (1988) 749. 
  
   Z.~Komargodski and A.~Schwimmer,
 ``On Renormalization Group Flows in Four Dimensions,''
  JHEP {\bf 1112} (2011) 099
  [arXiv:1107.3987 [hep-th]]. 
  
  H.~Casini and M.~Huerta,
  ``On the RG running of the entanglement entropy of a circle,''
  Phys.\ Rev.\ D {\bf 85}, 125016 (2012)
  doi:10.1103/PhysRevD.85.125016
  [arXiv:1202.5650 [hep-th]].
  
  \bibitem{Calabrese:2014yza} 
  P.~Calabrese, J.~Cardy and E.~Tonni,
  ``Finite temperature entanglement negativity in conformal field theory,''
  J.\ Phys.\ A {\bf 48}, no. 1, 015006 (2015)
  doi:10.1088/1751-8113/48/1/015006
  [arXiv:1408.3043 [cond-mat.stat-mech]].
  
\bibitem{Banks:1998dd} 
  T.~Banks, M.~R.~Douglas, G.~T.~Horowitz and E.~J.~Martinec,
  ``AdS dynamics from conformal field theory,''
  hep-th/9808016.
  
  \bibitem{Balasubramanian:1998de} 
  V.~Balasubramanian, P.~Kraus, A.~E.~Lawrence and S.~P.~Trivedi,
  ``Holographic probes of anti-de Sitter space-times,''
  Phys.\ Rev.\ D {\bf 59}, 104021 (1999)
  doi:10.1103/PhysRevD.59.104021
  [hep-th/9808017].
  
   \bibitem{Heemskerk:2010hk} 
  I.~Heemskerk and J.~Polchinski,
  ``Holographic and Wilsonian Renormalization Groups,''
  JHEP {\bf 1106}, 031 (2011)
  doi:10.1007/JHEP06(2011)031
  [arXiv:1010.1264 [hep-th]].
  
  T.~Faulkner, H.~Liu and M.~Rangamani,
  ``Integrating out geometry: Holographic Wilsonian RG and the membrane paradigm,''
  JHEP {\bf 1108}, 051 (2011)
  doi:10.1007/JHEP08(2011)051
  [arXiv:1010.4036 [hep-th]].
  
  D.~Elander, H.~Isono and G.~Mandal,
  ``Holographic Wilsonian flows and emergent fermions in extremal charged black holes,''
  JHEP {\bf 1111}, 155 (2011)
  doi:10.1007/JHEP11(2011)155
  [arXiv:1109.3366 [hep-th]].
    
  D.~Radicevic,
  ``Connecting the Holographic and Wilsonian Renormalization Groups,''
  JHEP {\bf 1112}, 023 (2011)
  doi:10.1007/JHEP12(2011)023
  [arXiv:1105.5825 [hep-th]].
  
  \bibitem{Ryu:2006bv} 
 S.~Ryu and T.~Takayanagi,
  ``Holographic derivation of entanglement entropy from AdS/CFT,''
  Phys.\ Rev.\ Lett.\  {\bf 96}, 181602 (2006)
  [hep-th/0603001]. 
  
  H.~Casini, M.~Huerta and R.~C.~Myers,
  ``Towards a derivation of holographic entanglement entropy,''
  JHEP {\bf 1105}, 036 (2011)
  doi:10.1007/JHEP05(2011)036
  [arXiv:1102.0440 [hep-th]].
  
  A.~Lewkowycz and J.~Maldacena,
  ``Generalized gravitational entropy,''
  JHEP {\bf 1308}, 090 (2013)
  doi:10.1007/JHEP08(2013)090
  [arXiv:1304.4926 [hep-th]].
    
   
   \bibitem{Jacobson:2003wv}
  T.~Jacobson and R.~Parentani,
  ``Horizon entropy,''
  Found.\ Phys.\  {\bf 33} (2003) 323
  [gr-qc/0302099].  
  
   \bibitem{Vidal:2002zz} 
  G.~Vidal and R.~F.~Werner,
  ``Computable measure of entanglement,''
  Phys.\ Rev.\ A {\bf 65}, 032314 (2002).
  doi:10.1103/PhysRevA.65.032314
  
 
    


\end{thebibliography}
\end{document}